# Near-unity light-matter interaction in mid-infrared van der Waals nanocavities

Haonan Ling,[1,†] Milad Nourbakhsh,[2,†] Vincent R. Whiteside,[2] Joseph G. Tischler,[2,*] Artur R. Davoyan[1,*]

[1]University of California, Los Angeles, [2]University of Oklahoma

[†]these authors contributed equally
[*]Correspondence to: tischler@ou.edu and davoyan@seas.ucla.edu

**Abstract:**
Accessing mid-infrared radiation is of great importance for a range of applications, including thermal imaging, sensing, and radiative cooling. Here, we study light interaction with hexagonal boron nitride nanocavities and reveal strong and tunable resonances across its hyperbolic transition. In addition to conventional phonon-polariton excitations, we demonstrate that the high refractive index of hexagonal boron nitride outside the Reststrahlen band allows enhanced light-matter interactions in deep subwavelength ($< \lambda/15$) nanostructures across a broad $7 - 8\ \mu m$ range. Near-unity absorption and high quality ($Q \geq 80$) resonance interaction in the vicinity of the transverse optical phonon is observed. Our study provides new avenues to design highly efficient and ultracompact structures for controlling mid-infrared radiation and accessing strong light-matter interaction.

**Main text:**

Mid-infrared (mid-IR) radiation plays an important role in a wide range of applications covering thermal imaging [1-3], mimicry [4, 5], molecular and bio-sensing [6, 7], waste heat management [8-12], and radiative cooling [13-15], amongst others. Many diverse pathways have been proposed and examined recently to control mid-IR radiation, including such systems as plasmon-polaritons in graphene and doped semiconductors [16-22], all-dielectric metasurfaces [23, 24], and polar dielectrics [25-32]. Of great interest are polar dielectrics, such as silicon carbide [27-30] and hexagonal boron nitride (hBN) [31-36], which owing to their intrinsically strong phonon resonances [30, 32] offer unique avenues for light-matter interactions in the mid-IR band. In addition, phonons can couple with optical modes forming highly confined hybridized phonon-polariton states [33, 37-40].

Physics of phonon-polaritons has received significant attention due to their unique hyperbolic dispersion [30, 32, 38] and ability to squeeze light well below the diffraction limit [28, 32]. As a result, deep subwavelength ($< 1\ \mu m$) phonon-polariton devices [28, 39] can be designed to control mid-IR radiation ($\lambda > 5\ \mu m$). A range of phonon-polariton systems, including waveguides [41], metasurfaces [29, 31, 42], and nanoscale resonators [27] have been proposed and studied for mid-IR radiation guidance, emission, and absorption. At the same time, phonon-polariton coupling exists in a relatively narrow wavelength range, inside the Reststrahlen bands (RB) [28, 32, 34], where the dielectric permittivity tensor $\bar{\bar{\varepsilon}}$ has negative values in one or two directions and the material exhibits hyperbolic dispersion [38]. The narrow range of existence of phonon-polariton states limits access to controlling a broader mid-IR spectrum and therefore, constrains possible applications.

Outside of the RB, the dielectric permittivity tensor $\bar{\bar{\varepsilon}}$ for both silicon carbide and hBN is positive for all components ($\bar{\bar{\varepsilon}} > 0$) [28, 32] and materials behave as highly dispersive dielectrics. In this case "all-dielectric" guided and resonant mid-IR optical modes can be excited [27, 29] in the bulk of the material. Importantly, owing to a very high quality of the phonon resonances, dielectric permittivity components can reach very high values ($> 100$) [28, 32]. In turn, high refractive index provides a pathway for enhanced light-matter interactions and design of nanoscale ($\ll \lambda$) systems. Light coupling to resonant silicon carbide nanostructures in the $\bar{\bar{\varepsilon}} > 0$ regime has been examined in [29], where excitation of Mie resonances was observed. In addition, use of high refractive index in hBN was proposed for low loss deeply subwavelength waveguides [43].

At the same time, while the study of phonon-polaritons has received broad interest, little attention has been paid to understanding and controlling "all-dielectric" modes in the vicinity of phonon resonances (i.e., in $\bar{\bar{\varepsilon}} > 0$ regime). One of the challenges in accessing this wavelength domain is associated with fabrication of subwavelength structures that preserve the nature of high-quality phonon excitations. Indeed, while phonon-polaritons can be easily excited at the surface of a bulk single crystal [40], resonant excitations in the $\bar{\bar{\varepsilon}} > 0$ regime necessitates careful design of the nanoscale structures. Layered van der Waals materials, such as hBN, provide a unique pathway to studying high quality (i.e., strongly resonant) light-phonon interactions down to a single atomically thin layer [34, 44].

Here, we study light-matter interactions within deeply subwavelength ($< \lambda/15$) hBN nanocavities in the vicinity of the transverse optical (TO) phonon resonance of the higher energy RB across the hBN hyperbolic phase transition. We demonstrate that such nanostructures allow tunable near-unity light absorption across a wide mid-IR band, including the region in which all permittivity components are positive (i.e., $\bar{\bar{\varepsilon}} > 0$). We investigate the dispersion of light-matter interaction across the upper RB and hyperbolic transitions as a function of nanocavity geometry. Excitation of multiple high-quality resonances ($Q$-factor exceeding 80) tunable through the hBN nanostructuring, and their interaction at the edge of the hyperbolic transition are demonstrated.

We begin our analysis by probing the resonant nature of light-phonon interaction in planar hBN films. For this purpose, we designed a planar nanoscale cavity, schematically shown in Fig. 1(a). Here the cavity is formed by a gold back reflector with an hBN slab of thickness $h$ placed atop of it. Owing to the high refractive index of hBN close to the TO phonon resonance at $\lambda_{TO} = 7.34\ \mu m$ [32], deeply subwavelength dielectric resonances are present in which the hBN film acts as a Fabry-Perot (FP) resonator when the condition $4\frac{n(\lambda)h}{\lambda} \simeq (2m + 1)$ is met (here $n(\lambda)$ is the real part of the in-plane hBN refractive index component, i.e., $n(\lambda) = Re(\sqrt{\varepsilon_\perp})$, $\varepsilon_\perp$ is the permittivity component along the hBN basal plane, perpendicular to the hBN crystal axis, and $m = 0, 1, ...$ is the resonance order).

The gold back reflector with its near perfect reflection in mid-IR further maximizes light interactions with hBN and stimulates critical coupling [45-49]. We note that a similar geometry was explored in a number of works studying light interaction with high index materials [48, 49]. Further, to minimize potential excitation and coupling with gold surface plasmon modes, as well as assist with cavity fabrication (Methods), we introduce a 100 nm silicon spacer layer between the gold and hBN (Fig. 1(a)). The relatively low refractive index of silicon ($\varepsilon_{Si} \simeq 11.69$ [50]), as compared to that of hBN near its TO resonance, and small spacer thickness ensure that the dispersion of modes excited in hBN is just perturbed weakly by the added silicon layer. The nanocavity geometry allows performing mid-IR reflectivity, where resonances correspond to the excitation of hBN film eigen-modes (note that, as absorption in gold and silicon is negligibly small, only excitations of hBN are seen in the reflection spectra).

We prepared a range of such hBN nanocavities with $h$ varying from 50 nm to 485 nm as determined by atomic force microscopy (AFM). Specifically, single crystal hBN flakes were mechanically exfoliated atop a silicon/gold substrate (Fig. 1(a) inset); details of nanocavity fabrication are provided in the Methods section. We then studied the fabricated nanocavities through reflectivity measurements in the range of 6.5 $\mu m$ to 8.5 $\mu m$ with the use of Fourier Transform Infrared (FTIR) microscopy (Methods). Fig. 1(b) shows experimental reflectance spectra for several fabricated cavities of different hBN film thickness. We observe sharp resonances in reflection for wavelengths larger than $\lambda_0 \simeq 7.295\ \mu m$, i.e., in the range where all permittivity tensor components are positive including the perpendicular ones ($\varepsilon_\perp > 0$) and hBN behaves as a highly anisotropic uniaxial dielectric crystal.

Evidently, the resonance energy depends on the hBN nanocavity thickness. For a very thin hBN film ($h = 50\ nm$) the resonance is observed at $\lambda = 7.387\ \mu m$. As the hBN thickness increases the resonance shifts to longer wavelengths, as expected for a FP cavity with normal dispersion of refractive index dispersion, $n(\lambda)$. Notably, for the 200 $nm$ thick hBN we observe $\approx$ 90% absorption at $\lambda \simeq 7.58\ \mu m$. Such strong absorption at this thickness can be attributed to the critical coupling condition (which is also evident from simulations shown in Fig. 1(c)) [46-48]. The resonance for a 200 $nm$ thick film exhibits a relatively high-quality factor, $Q \simeq 40$, which is important for the design of narrow band infrared thermal emitters and filters [1, 11, 51, 52]. As the film thickness increases ($h > 300\ nm$), a second order FP resonance emerges.

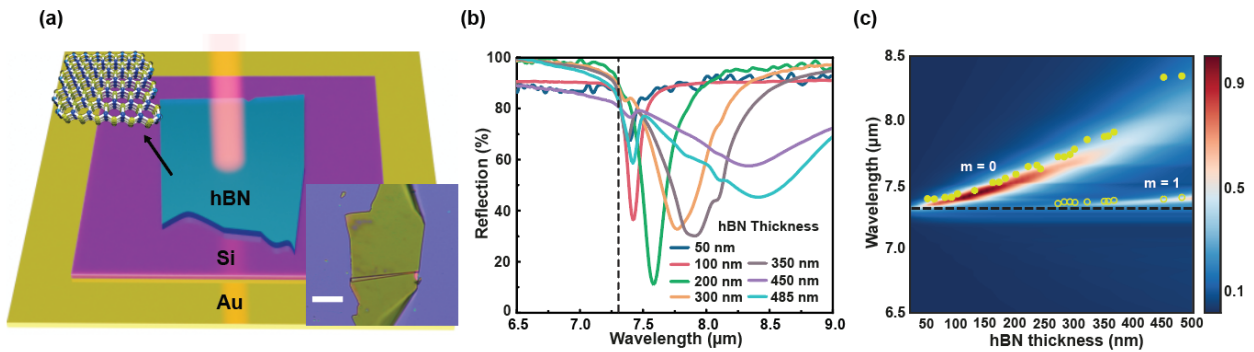

FIG. 1. hBN thin film nanocavity. (a) Schematic illustration of the nanocavity design. Inset shows an optical micrograph of an as-exfoliated hBN flake atop of a 100 nm Si spacer layer deposited on 100 nm thick Au substrate. Scale bar: 20 μm. (b) Measured FTIR reflection spectra of multilayer structures shown in (a) for different hBN flake thicknesses. (c) Color map of the calculated absorption spectra in the nanocavity as a function of the hBN thickness. Yellow symbols indicate positions of the resonances measured for different hBN flake thicknesses. The dashed line in (b) and (c) indicates the onset of RB near the transverse optical phonon energy, at which $\varepsilon_\perp$ changes its sign.

To better understand the observed dynamics, we compare experimental results with numerical simulations. In our simulations we study plane wave reflection under normal incidence. A respective color map of absorption as a function of excitation wavelength and hBN thickness is shown in Fig. 1(c). For convenience we overlay the experimental absorption resonance maxima

(i.e., reflection minima in Fig. 1(b)) atop the simulated color map. A very good agreement between numerical simulations and experimental measurements is observed (despite that our simulations are for normal incidence only and the experiments are performed with average incidence angle of 15°).

Simulations provide a more detailed insight into light-matter interactions in hBN nanocavities. Firstly, we observe two distinct absorption resonance branches emerging right near $\lambda_0 = 7.295 \, \mu m$, i.e., near the onset of the hyperbolic transition in hBN around TO phonon resonance (note that hyperbolic transition occurs slightly below the TO phonon resonance, i.e., $\lambda_0 < \lambda_{TO}$). These branches correspond to the first two FP cavity modes ($m = 0$ and $m = 1$) and exhibit a cut-off with the hBN thickness. The thickness cut-off condition can be understood from the FP resonance condition, $4 \frac{n(\lambda)h}{\lambda} \simeq (2m + 1)$: $h_{cut} \propto \frac{2m+1}{n_{max}}$, where $n_{max} = \max(n(\lambda))$ is the maximum of the hBN in-plane refractive index reached in the vicinity of the hyperbolic transition at $\lambda_0$ [32]. For the structure studied here we obtain $h_{cut} \simeq 50 \, nm$ for the $m = 0$ and $h_{cut} \simeq 270 \, nm$ for the $m = 1$ modes respectively. In principle, as the hBN thickness increases further we expect that higher order nanocavity resonance will become visible.

In addition to the cut-off, we also observe near-unity light absorption at $h \sim 150 \, nm$ for the fundamental nanocavity resonance (i.e., $m = 0$ mode). Such near-unity absorption can be understood as reaching conditions close to the critical coupling, at which the rate of light absorption in the nanocavity exactly matches the rate of incoming flux [53]. Observed near-unity light absorption is also associated with a narrow bandwidth, which we attribute to a strong dispersion of the hBN permittivity near the TO resonance [32]. In the range where $\varepsilon_\perp < 0$ (i.e., in the regime of hyperbolic dispersion inside the RB), we observe slight absorption right at the edge of the RB, as was also recently investigated [34]. Deeper into the RB the permittivity reaches high negative values thus light does not penetrate through the material. Despite its simplicity, the planar nanocavity studied here already reveals a very rich dynamics and demonstrates that structures as thin as a few hundred nanometers (i.e., $< \lambda/30$) can allow access to a broad range of infrared wavelengths outside of the RB.

Next, to examine the way in which the dispersion can be further engineered, we extend our study to patterned hBN disk arrays, shown in Fig. 2(a). While planar cavity studied in Fig. 1(a) already exhibits very rich physics, patterning it into nanostructures allows accessing a broader set of phenomena at both sides of the hyperbolic transition at $\lambda_0$. A range of patterned hBN nanostructures have been studies previously with a focus on phonon-polariton excitations in $\varepsilon_\perp < 0$ regime, including 1D gratings [31], bowtie structures [54], pillars [55], photonic crystal slabs [56, 57], and disk resonators [32]. However, here our objective is to access modes in the "all-dielectric" range of the hBN permittivity (i.e., outside of the RB where $\varepsilon_\perp > 0$) and understand the evolution of light-matter interactions across the hyperbolic transition around the TO resonance. To understand the nature of such new states of light—hBN interactions, we consider two limiting cases: ultrathin film with $h \simeq 100 \, nm$ and a thicker film with $h \simeq 500 \, nm$, in which a broader set of excitations – especially when patterned – are expected.

Fig. 2(a) shows a micrograph of a $97 \, nm$ thick hBN film patterned by electron beam lithography into four disk arrays of different radii atop of the Si/Au substrate (i.e., 100 nm Si spacer on top of a gold back reflector) (see Methods for details on fabrication and characterization procedures). A respective AFM cross-section is shown in the inset of Fig. 2(a). In Fig. 2(b) we measured the fabricated disk arrays through scanning electron microscopy (SEM). The arrays were designed with a period of 5 μm and radii $r = 0.8$ μm, 1.3 μm, 1.6 μm, 1.9 μm, respectively (Fig. 2(b)). Overall 6 to 9 disks were fabricated per given array since our objective is to study arrays of the same thickness. Therefore, we fabricate all four arrays on the same exfoliated flake. As a result, the useful area size is limited by the quality of the exfoliated flake, which precludes fabricating large area arrays.

Measured FTIR reflection spectra for all four fabricated arrays are plotted in Fig. 2(c). For all four arrays we observe only one absorption resonance in the $\varepsilon_\perp > 0$ regime with $Q$-factors ranging from $50 - 61$ (see inset in Fig. 2(b)). This resonance exhibits a weak dependence on the disk radius. We attribute the nature of this excitation to a FP mode, similar to that observed in an unpatterned hBN nanocavity (Figs. 1(a)-1(c)). For a 97 $nm$ thick hBN film the resonance for an unpatterned film occurs at $\lambda = 7.385$ $\mu m$ (Fig. 1(c)), which corresponds well with the measurements in Fig. 2(c).

A related numerical simulation of absorption spectra as a function of the hBN disk radius is plotted in Fig. 2(d). In our simulation we assume infinite periodic structures excited by a plane wave under normal incidence (i.e., a rough approximation of our experiment in which a few element-array is excited by a Cassegrain objective yielding an average oblique incidence of $15^o$). For the sake of comparison, the positions of the measured resonances are overlayed atop the calculated color map in Fig. 2(c). Despite the normal incidence approximation in our simulations, a great agreement between simulations and measurements is seen. Outside the RB where the in-plane hBN permittivity is positive (i.e., $\varepsilon_\perp > 0$), we find that there is only one absorption resonance, which varies weakly with the disk radius. In the limit of large disk radius, the patterned structure behaves as a continuous film and the resonance matches that of a planar nanocavity for a given thickness (see Fig. 1(c)). On the other hand, as the disk size shrinks the strength of the resonance decreases and a slight broadening of the resonances is observed. Overall, for disk radii $<$ 0.6 $\mu m$ resonances vanish and cannot be seen in our simulations as expected.

The observed dynamics can be understood via an effective medium picture: as the disk size is reduced the effective permittivity of the overall hBN film is also reduced resulting in the decrease of the resonance frequency; at some point, as the film thickness is fixed, a cut-off condition is met (see also discussion pertaining to Fig. 1(c)). Numerical simulations also predict multiple resonance branches inside the RB, which are familiar phonon-polariton modes seen in a number of experiments [30, 32, 33]. The excitation of these modes is not clearly observed in our experiments. Taller hBN pillars are needed in order to observe phonon polariton modes [39, 40], as we show below for our 500 nm flake in Fig. 3. Even in the simulations the strength of these modes, as seen through absorption resonances is rather weak (as compared with Fig. 3(c)).

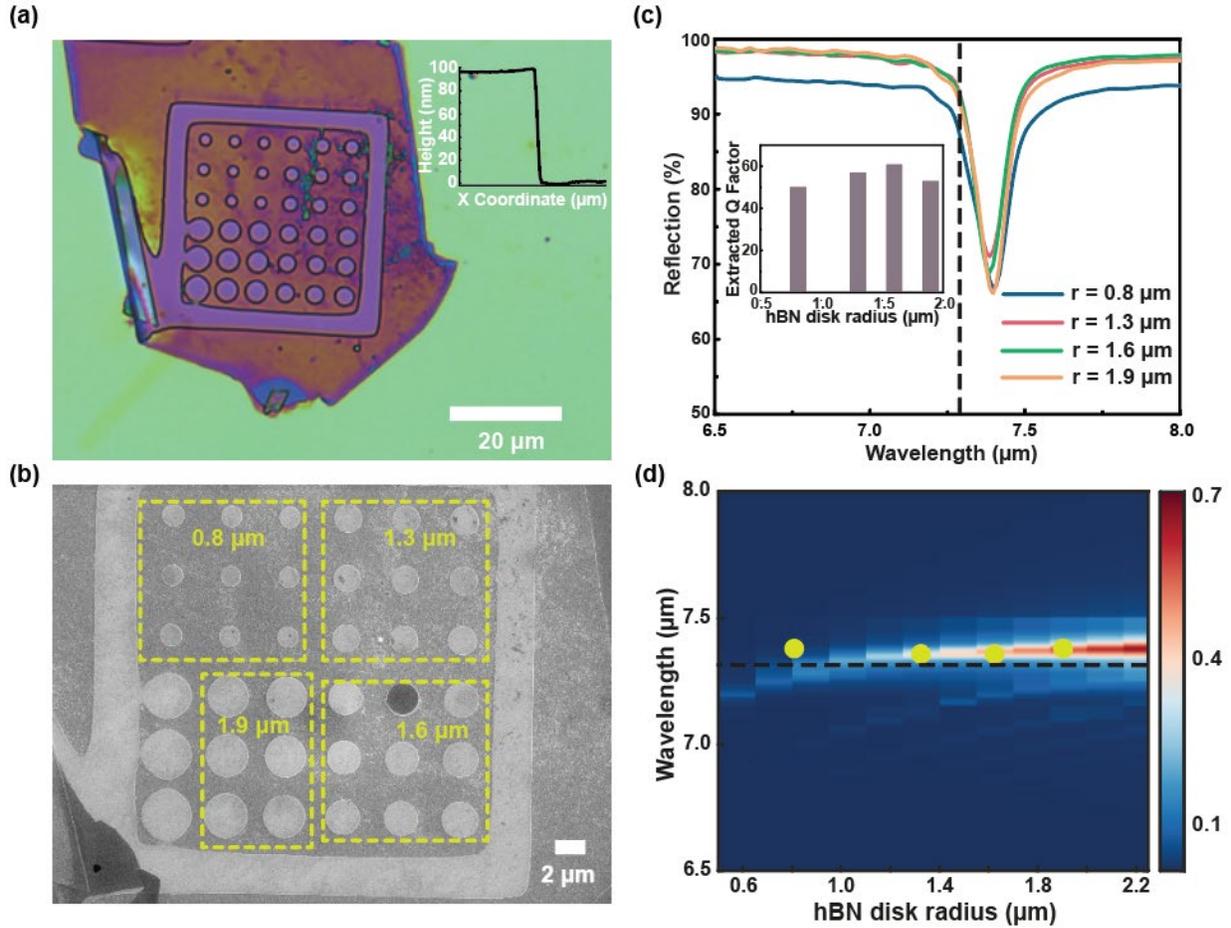

FIG. 2. Thin film hBN disk array study. (a) Optical micrograph of fabricated hBN disk arrays of 4 different radii in a 97 nm thick film. Inset shows thickness profile of the as-exfoliated flake measured by AFM. (b) SEM image of disk arrays shown in (a) with different radii marked. (c) Measured FTIR reflection spectra for 4 fabricated hBN disk arrays of different radii. Inset shows corresponding extracted Q-factors. (d) Color map of the simulated absorption spectra as a function of the hBN disk radius (5 μm period is assumed). Positions of the measured resonances are marked for convenience. The dashed line in (c) and (d) indicates the onset of RB near the transverse optical phonon energy, at which $\varepsilon_\perp$ changes its sign.

Moving forward, we study a thicker hBN film of $h \simeq 500\ nm$, in which case we anticipate a wider range of excitations. Following the same procedure, we pattern an exfoliated 500 nm flake into 7 disk arrays with periodicity of 5 μm and radii r = 0.8 μm, 1.1 μm, 1.35 μm, 1.5 μm, 1.6 μm, 1.8 μm, and 2.0 μm, respectively. Fig. 3(a) shows a respective micrograph of a part of the flake and AFM profile. Corresponding reflectance spectra of disk arrays measured by FTIR spectroscopy are plotted in Fig. 3(b). A set of distinct resonances are observed in reflectivity, which correspond to rich dynamics in the patterned thick hBN resonators. Firstly, unlike the 97 nm case studied earlier, for a thicker film we observe a set of absorption resonances inside the RB corresponding to phonon-polariton modes. We also observe that for a number of structures, resonances appear to be "pinned" close to $\lambda_0 = 7.295\ \mu m$ (i.e., the edge of the RB where

$\varepsilon_\perp$ changes its sign). Outside the RB, where $\varepsilon_\perp > 0$ we find a set of resonances that strongly depend on disk radius as well.

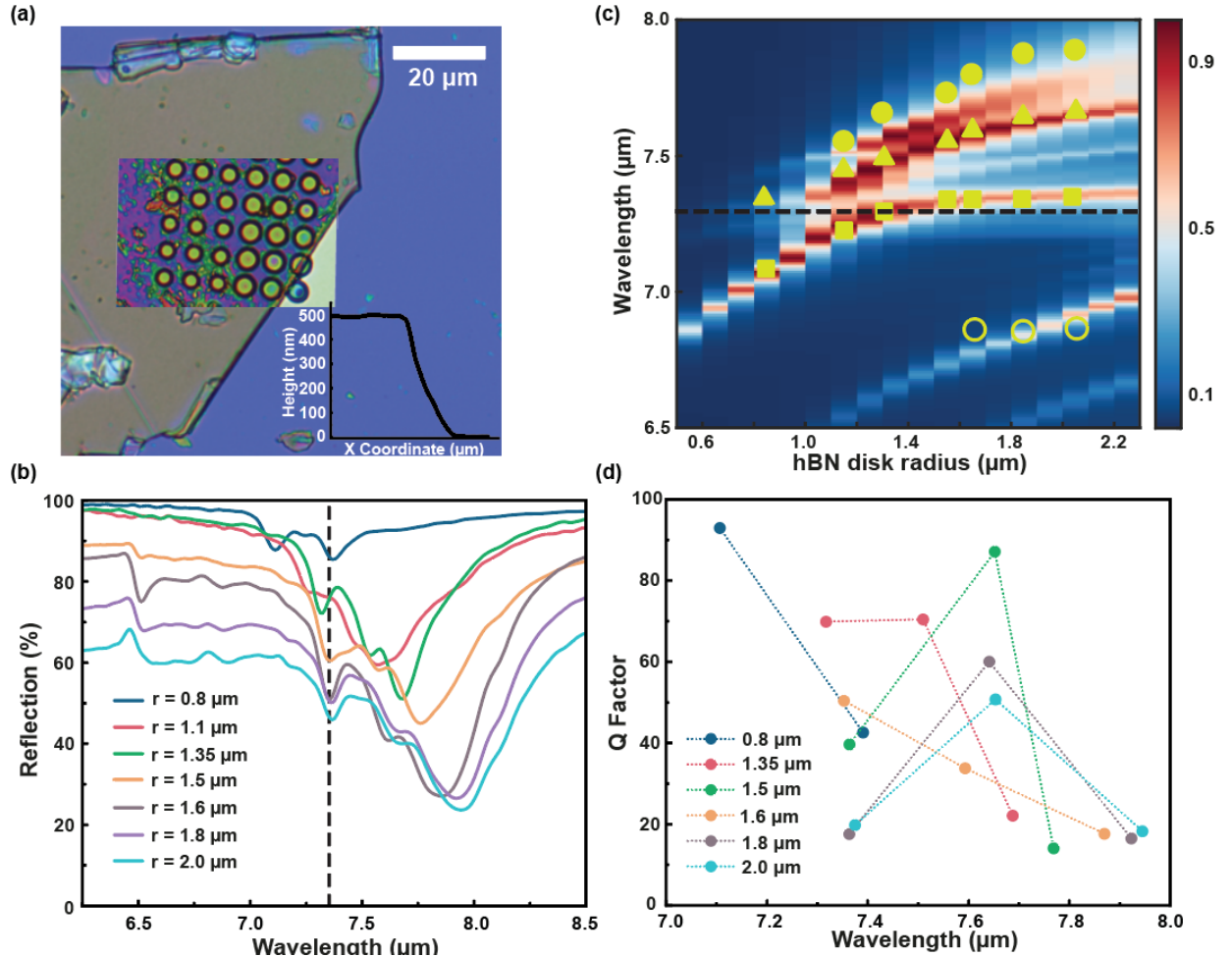

FIG. 3. Evidence of strong light-matter interactions in hBN disk arrays. (a) Optical micrograph of the as-exfoliated 500 nm hBN flake with an overlaid micrograph of fabricated disk arrays. Inset: thickness profile of the as-exfoliated flake measured by AFM. (b) Measured reflection spectra for fabricated hBN disk arrays of 7 different radii. (c) Simulated absorption spectra of hBN disk arrays as a function of hBN radius. The period is 5 μm and the hBN thickness is 500 nm. The markers denote positions of resonances extracted from the measured reflectance spectra in panel (b). The dashed line in (b) and (c) indicates the onset of RB near the transverse optical phonon energy, at which $\varepsilon_\perp$ changes its sign. (d) Extracted Q-factors for measured resonances in the vicinity of the TO phonon as a function of the resonance wavelength.

To better understand the dynamics observed in our measurements, we performed numerical simulations. In Fig. 3(c) we plot the simulated absorption spectra as a function of the hBN disk radius assuming $h = 500\ nm$. Firstly, we note an excellent agreement between experimental results and numerical calculations. In these simulations three distinct phonon-polariton branches are seen in the $\varepsilon_\perp < 0$ regime. Strong dispersion of these branches with the disk radius is evident. Such behavior can be explained by the dispersion of phonon-polaritons in planar hBN films [58] and by the hyperbolic nature of the optical dispersion inside the hBN RB (i.e., $\varepsilon_\parallel > 0$ and $\varepsilon_\perp < 0$,

$\frac{k_\parallel^2}{\varepsilon_\perp} + \frac{k_\perp^2}{\varepsilon_\parallel} = \frac{\omega^2}{c^2}$, where $\omega$ is the frequency and $c$ is the speed of light) [32]. Specifically, disk size, $r$, dictates the effective wavevector of the excited phonon-polariton $k_{pp} \propto \frac{1}{r}$, which has predominant in-plane orientation, i.e., $k_\perp \sim k_{pp}$.

As a result, smaller disk radii imply that modes with a larger in-plane wavevector, $k_\perp$, are excited. From the dispersion relation it is evident that such modes are excited away from the $\lambda_0 = 7.295\ \mu m$ deep into the RB when $|\varepsilon_\perp| \to 0$ (i.e., for a small negative in-plane permittivity which occurs deep into the RB). On the contrary, larger disk sizes require smaller $k_{pp}$, which is accessed at larger $|\varepsilon_\perp|$. At the same time, it is known that for relatively thick films ($h > 400\ nm$) already with moderate values of $k_{pp}$ phonon resonances saturate at around $\lambda_0$ [58, 59]. Such "pinning" is seen in our simulations and experiments for the fundamental phonon-polariton mode at $r > 1.4\ \mu m$. Larger disk radii also suggests excitation of higher order phonon-polariton modes. Their dispersion is similar to that of the fundamental modes, i.e., red-shifts with increasing disk radius.

Outside the RB where $\varepsilon_\perp > 0$, i.e., for $\lambda > \lambda_0$ several resonant branches are distinctly observed. We note that for $h = 500\ nm$ a simple planar nanocavity supports two FP resonances (at $\lambda = 7.39\ \mu m$ and $\lambda = 8.02\ \mu m$, respectively, see Fig. 1(c)). When patterned into a disk array these FP resonances are augmented with Mie-like modes supported within individual disks. As a result, a very complex picture of absorption is seen for $\varepsilon_\perp > 0$. Nevertheless, some intuitive understanding on the observed evolution can be made. FP modes inherit their features from an unpatterned cavity and can be understood from the effective medium picture (i.e., are defined by the condition $\frac{n_{eff}(\lambda)h}{\lambda} = const$, where $n_{eff}(\lambda)$ is the effective refractive index of patterned hBN film). Patterning of the hBN film "dilutes" the hBN permittivity and results in the decrease of the effective refractive index leading to a resonance blue shift (a behavior evidently seen in Fig. 2(d)).

On the contrary, Mie-like disk resonances, are mainly defined by the disk size, and are excited when the condition $r \propto \frac{\lambda}{2n(\lambda)}$ is met. Smaller disk size, $r$, necessitates a higher hBN refractive index, $n(\lambda)$, to support the mode. Higher index, in turn, is observed closer to the hyperbolic transition at $\lambda_0$. Notably, simulations predict >90% absorption which can be tuned across a wide wavelength range for several of the observed branches.

In Fig. 3(d) we examine quality factors for the experimentally observed resonances. For this purpose, we fit measured spectra with a sum of Lorentz oscillators (Methods). Such fitting reveals that resonances have high quality factors ($Q$-factors in the range $20 - 90$) and can be controlled with the structure geometry. For example, for $r = 0.8\ \mu m$ disk array we obtain $Q \geq 90$ in the $\varepsilon_\perp < 0$ regime. Similarly high quality factor is observed for $r = 1.5\ \mu m$ disks in $\varepsilon_\perp > 0$ regime ($Q \simeq 90$ at $\lambda \simeq 7.6\ \mu m$).

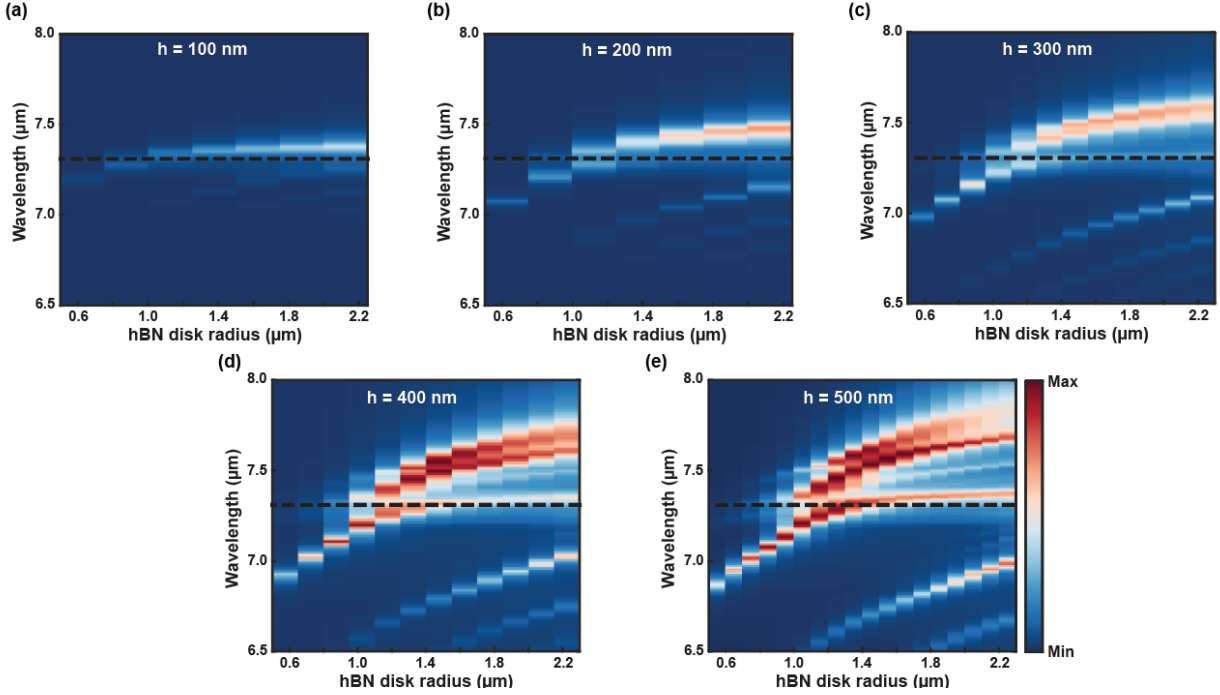

FIG. 4. Wideband tunable strong mid-IR light-materials interaction. Calculated absorption spectra in hBN disk arrays with the disk radius for (a) h=100 nm, (b) h = 200 nm, (c) h = 300 nm, (d) h = 400 nm, and (e) h = 500 nm films, respectively. Dashed lines in each plot indicates the onset of RB near the transverse optical phonon energy, at which $\varepsilon_\perp$ changes its sign.

Finally, we investigate numerically the evolution of light absorption at the edge of the RB, around the hyperbolic transition, as a function of the hBN film thickness. Figs. 4(a) – 4(e) show calculated color maps of the absorbance spectra of hBN disk arrays as a function of the hBN disk radius for different hBN thickness. For thin films (100 $nm$ – 300 $nm$, Figs. 4(a) – 4(c)) we observe only one resonance mode in the $\varepsilon_\perp > 0$ regime. As the hBN film thickness grows this mode shifts to longer wavelengths, as is expected for a mode inheriting its origin from FP nanocavity studied in Fig. 1(c). For thicker films higher order modes emerge and disks are now capable of sustaining Mie-like resonances. "Pinning" of the phonon polariton branches to the edge of the RB with thickness increase is also evidently seen and resembles dispersion of photon-polaritons in thin hBN films [58, 59]. The strongly dispersive nature of the resonances with the structure dimensions enables engineering of the absorptivity and emissivity over a wide range of mid-IR frequencies, including outside of the RB. Hence, strong and tunable light absorption from $\lambda = 7\ \mu m$ to $\lambda = 8\ \mu m$ in deeply subwavelength ($< \lambda/15$) structured films is possible.

In conclusion, our study reveals that hBN, owing to its high-quality phonon excitations, enables enhanced of light-matter interactions across an ultrawide mid-IR band spanning both sides of the hyperbolic dispersion (i.e., inside RB where $\varepsilon_\perp < 0$ and outside of it where $\varepsilon_\perp > 0$). The overall evolution of the resonances in $\varepsilon_\perp > 0$ and $\varepsilon_\perp < 0$ regimes may also be viewed as a manifestation of avoided crossing due to a strong coupling of the light to phonon excitations in hBN nanostructures [60, 61]. In this case hyperbolic dielectric permittivity of hBN is understood via intrinsic strong coupling of vacuum fields with phonons coupling [61], whereas modes in the $\varepsilon_\perp > 0$ regime as "upper polaritons" and phonon polaritons in $\varepsilon_\perp < 0$ regime as "lower polariton" branches, respectively. Such "avoided crossing" feature right at the onset of the hyperbolic transition is seen for a 300 $nm$ hBN film (Fig. 4(c)).

Our nanocavity design may therefore serve as a platform to study strong coupling of infrared light with phonons. Furthermore, in addition to the commonly studied phonon-polariton branches, our work demonstrates that high refractive index of hBN and its extreme dispersion in $\varepsilon_\perp > 0$ range enable new approaches to harnessing mid-IR radiation in deep subwavelength ($< \lambda/15$) structures. Narrow band resonances with quality factors as high as 90 featuring near-unity absorption across $7\ \mu m - 8\ \mu m$ band can pave the wave to infrared light sources [11], thermal management [8], and spectroscopy [7].

**Acknowledgments**
J.G.T. acknowledges support by National Science Foundation under Grant OISE-2230706.
A.R.D. acknowledges support by DARPA Grant # HR00112320021.

**Author contributions**
H.L. designed and fabricated samples, performed numerical simulations and derivations. M.N. and V.R.W. performed FTIR measurements. All authors contributed to discussion, analyzing results and manuscript writing. J.G.T. and A.R.D initiated and supervised the project.

**Methods**

Sample fabrication

Samples were fabricated atop of a Si wafer. Around 100 nm gold layer was first deposited atop of a Si wafer using CHA Mark 40-1 e-beam evaporator with a thin titanium adhesion layer (Au:Ti = 100:10 nm). Subsequently around 100 nm polycrystalline Si layer was deposited using CHA Solution e-beam evaporator with a thin titanium adhesion layer (Si:Ti = 100:10 nm).

hBN flakes were mechanically exfoliated from bulk crystal (2D Semiconductors) using a scotch tape onto the prepared substrates. The topography and thickness of exfoliated flakes is examined with optical microscopy (thickness of desirable flakes is predetermined based on color) and AFM (Bruker Dimension FastScan with ScanAsyst).

To pattern hBN flakes into disks, a thin layer of tungsten about 20 nm (serves as a charge dissipation layer and an adhesion layer) was deposited with a sputter coating (Denton Discovery 550 Sputter). Next, negative e-beam resist (MaN 2403) was spin-coated on top of the chip with 3000 rpm for 30 seconds. The resultant resist thickness is around 300 nm. The chip was baked at 95 °C for 1 minute. Then the flake was patterned by e-beam lithography (Raith EBPG 5000+ES) with 100 keV beam energy. After that, the written flake was developed with MF-319 solution followed by 90 seconds rinse in DI water. The resist exposed by e-beam remained on the flake after development, serving as an etching mask as well. Afterwards fluorine reactive ion etching (RIE) (Oxford PlasmaLab 80+) was used to remove both tungsten (recipe: 40 sccm $SF_6$, forward power of 70 W, and pressure at 30 mtorr for 25 seconds) and exposed hBN (recipe: 40 sccm $CHF_3$, 4 sccm $O_2$, forward power of 55 W and pressure at 40 mtorr). After the exposed hBN has been etched away, the remaining resist was removed by immersion in hot acetone or oxygen plasma, and the tungsten layer washed away with a 30% $H_2O_2$ solution.

Optical simulations

For results on unpatterned hBN multilayer cavity shown in Figure 1, the solutions were obtained with a transfer matrix method. Wave Optics module from COMSOL Multiphysics software was used to solve for reflection and absorption spectra of hBN disk arrays shown in Figures 2-4.

Fourier transform infrared (FTIR) spectroscopy

The reflection measurements were performed with a Bruker Hyperion 2000 infrared microscope connected to a Bruker Tensor 27 FTIR spectrometer. The infrared microscope is equipped with a mercury-cadmium-telluride (MCT) detector and a 15x reflective objective, providing a working distance and numerical aperture of 23.7 mm and 0.4, respectively. The Tensor 27 spectrometer, equipped with a standard KBr beam splitter, covered a spectral range from 370 cm$^{-1}$ to 7500 cm$^{-1}$. All measurements were carried out under room temperature and ambient pressure condition, with nitrogen gas purging to remove moisture. The sample stage is positioned perpendicular to the light beam, and the samples were exposed to unpolarized light at an incident angle ranging from 9.8° to 23.6°. The reflection spectra were recorded in the wavenumber range of 1000 cm$^{-1}$ to 2000 cm$^{-1}$ (corresponding to a wavelength range of 5 μm to 10 μm) with a scan number of 60 and spectral resolution of 4 cm$^{-1}$.

Data Analysis

The absorption curves corresponding to experimental data in Figure 2c has been fitted using Gaussian function of $f(x) = f_0 + Ae^{-4ln(2)(x-x_c)^2/w^2}/w\sqrt{\pi/4ln(2)}$. Measurements in Figure 3b were fitter by a sum of Lorentz oscillators: $f(x) = \sum y_{0i} + 2A_i/\pi + w_i/(4(x-x_{c_i})^2 + w_i^2)$. In these functions $x_{c_i}$ corresponds to observed resonance frequencies, and $y_{0i}$, $w_i$, and $A_i$ are fitting parameters.